# α-Ta (111) Thin Films for Qubit Applications: A Study of Thickness Dependence and Universal Scaling

Nate Price[1], Jose Gutiérrez[2], Sushant Padhye[3], Sara McGinnis[1], Carter Wade[1], Huma Yusuf[4], Lakshan Don Manuwelge Don[1], Kurt Eyink[5], J. Guerrero-Sánchez[6], Evgeny Mikheev[4], Joseph P. Corbett[1]

1. Miami University, Department of Physics, Oxford, OH 45056, USA
2. Centro de Investigación Científica y de Educación Superior de Ensenada, Baja California, Código 22860, México
3. Department of Electrical and Computer Engineering, University of Cincinnati, Cincinnati, OH 45219, USA
4. Department of Physics, University of Cincinnati, Cincinnati, OH 45211, USA
5. Materials and Manufacturing Directorate, Air Force Research Laboratory, Wright Patterson Air Force Base 45433
6. Centro de Nanociencias y Nanotecnología, Universidad Nacional Autónoma de México, Apartado Postal 14, Ensenada Baja California, Código Postal 22800

**Abstract**

We explore the growth of α-Ta thin films ranging from ultra-thin (~2 nm) to thick (~250 nm) films grown by sputter epitaxy on c-plane sapphire substrates. We utilized 100 W power with a 32 mTorr sputter pressure at 650 °C substrate deposition temperature. We used X-ray diffractometry to extract the lattice constant and growth orientation of the films, finding a mono-oriented (111) films with a lattice spacing in agreement with a bulk Ta value of 3.31 Å. X-ray reflectometry is used to characterize the native oxide, film, and substrate-film interface as a function of thickness. We observe a very smooth morphology with an average roughness of ~700 pm, as determined by both reflectometry and atomic force microscopy. The film nucleates with small islands, whose terrace width grows linearly as a function of thickness until ~150 nm, where the terrace width becomes constant. From our reflectometry measurements, we uncover a pseudomorphic layer with a critical thickness of ~1 nm and a self-limiting amorphous oxide that grows to a thickness of ~2.25 nm; both of these layers are independent of thickness beyond ~4 nm total thickness.

We also studied the superconducting transition through electronic transport measurements using the Van der Pauw method to measure resistivity as a function of temperature. We observed a smooth evolution in critical superconducting temperature with total film thickness, from ~2.9 °K for 7.5 nm to ~4.2 °K for 269.2 nm, as expected from universal thickness scaling in superconductors.

Density functional theory simulations were used to understand the oxidation process at the top surface layers of α-Ta (111). We observed that as the oxygen content on the surface increases, the Ta progressively loses its crystalline structure. Significant structural distortions occur when the Ta:O ratio exceeds 1:1, forming an amorphous TaO phase. We find the simulated oxide density to match well to the measured density range from reflectometry. Likewise, the structural variation of the amorphous oxide is in a range of ~0.8 nm, in good agreement with the reflectometry measurements finding, on average, a variation of ~0.7 nm.

**Introduction**

Quantum device performance is typically limited by disorder originating from the material synthesis procedure and the subsequent lithographic patterning[1–24]. Practically, the superconducting circuit technology space is dominated by a remarkably small number of materials: sapphire or high-resistivity silicon as low-loss insulating substrates with Al[4, 5, 16, 25–37] and Nb[17, 23, 38–50]as patterned superconductor films. There is substantial evidence for the importance of dielectric dissipation and fluctuations at accidental two-level systems (TLS), predominantly at interfaces and surfaces, encouraging the development of cleaner and more ordered insulator/superconductor material systems[1–6, 25, 38, 49, 51–57]. An exciting recent success story is the fabrication of qubits with α-Tantalum as the superconductor, leading to improved coherence times and large q-factors[56, 58, 58, 59, 59–62]. This improvement was believed to be linked to a more favorable chemistry of oxygen accommodation in Ta[60, 63, 64].

Deposition of Ta films is limited to a few methods due to the high melting point and lack of aqueous solutions suitable for electrochemical methods. Electron-beam and sputtering PVD are best suited to prepare Ta films[65–83]. Ta grows with two polymorphs, an α-phase, which is body-centered cubic with a lattice constant of 0.33029 nm, and β-phase, which is tetragonal and has lattice constants of 1.0194 nm and 0.5313 nm for the c-axis. The α-phase is best suited for SCRs with a transition temperature of ~4.3 K and lower resistivities compared to the metastable β-phase[56, 58, 68, 73, 76, 80, 82]. Residual oxygen content while sputtering is known to affect the phase of the material, where appreciable background oxygen favors a β-phase formation, although the exact critical concentration has not been determined. Dominating the recent α-Ta success in qubit research are (110) oriented α-Ta films on sapphire with up to 0.5 millisecond decoherence time[24, 58, 59, 63, 64, 84, 84–88]. Recently, α-Ta (111) grown films emerged with outstanding Q-factors up to 2 million even when not necessarily epitaxial.[37].

In this work, we study α-Ta (111) thin films as a function of film thickness to understand film growth, densification, surface oxidation, surface roughness, and universal scaling of the critical temperature[89]. The α-Ta films nucleate and grow with a (111) orientation under our ultra-high vacuum conditions and at 650 C on sapphire (001) or hexagonal $Al_2O_3$. A self-terminating amorphous oxide of ~2 nm in thickness forms upon exposure to residual oxygen and then air, after which it becomes a stable thickness. An interfacial layer between the $Al_2O_3$ substrate and α-Ta forms at ~0.9 nm thickness with a density close to that of bulk Ta. The resultant Ta (111) oriented film grows with a smooth surface roughness of ~ 500 pm and a density of ~70% of bulk α-Ta. The terrace width of the films grows continuously with film thickness until ~150 nm thickness, where the terrace width stabilizes.

**Experiment**

All films were grown with a custom-built ultra-high vacuum sputter epitaxy system with a base pressure of $1\times10^{-10}$ Torr. A 99.99% purity 1.3" diameter Tantalum target served as the source material purchased from Plasma Materials Inc. Deposition occurred at a fixed direct current (DC) power of 100 W under a 32 mTorr Ar pressure. A residual gas analyzer from Extorr Inc. provided a detailed mass spectrum of the residual chamber gasses up to 110 amu. Residual reactive gasses were carefully monitored prior to deposition and while the substrate was at elevated temperatures up to 1200 C to guarantee all oxygenating species were below a $1\times10^{-9}$ Torr threshold. Ta flux calibrations were determined using an atomic force microscope as a profilometer of shadow-masked Ta film grown. Atomic force imaging was performed on a commercial multimodal high-resolution AFM provided by AFMWorkshop. X-ray diffraction measurements were performed utilizing a Rigaku SmartLab for X-ray reflectometry (XRR) or a Bruker D8 Advance for symmetric scans.

Transport measurements were performed in a pulse-tube cryostat (Lakeshore Shi-4-2) from room temperature to a base temperature of 2.9 K. An on-chip thermometer (Lakeshore Cernox 1050 HT) was bonded to the sample holder and used to measure temperature. Van der Pauw's resistance of the films was measured on cooldown, stabilizing constant temperature in steps 0.1 K. Keithley 2400 was used for sourcing DC current (swept in 101 steps between ±100 µA) and Keithley DAQ6510 for DC voltage measurement. Two orthogonal van der Pauw orientations were measured in adjacent separate cooldowns from 5 K to base temperature.

**Computation**

The $\alpha$-Ta (111) surface structure was investigated using Quantum mechanical calculations within the Density Functional Theory (DFT) framework as implemented in the Vienna Ab Initio Simulation Package (VASP) code [90–92]. The electron-ion interactions were simulated using the Projector Augmented Wave (PAW) method [93], with an optimized energy cutoff of 400 eV. The exchange-correlation energy is treated according to the generalized gradient approximation (GGA) with the Perdew-Burke-Ernzerhof (PBE) parametrization [94]. The body-centered cubic structure of Ta was fully optimized, and it has a lattice parameter of 3.3 Å, which agrees with the experimental one. Then, it was projected along the [111] direction to design the surface. The α-Ta (111) surface was simulated using the supercell method, considering a 2 × 2 periodicity slab. It has a thickness of 13 atomic layers and ~23 Å of vacuum gap to preclude undesirable interactions. The oxidation process was carried out in stages, gradually increasing the oxygen content in the top three surface layers. In each stage, the coordinates of all atoms were relaxed without any constraint. The Brillouin zone was sampled using a k-points grid of 3 × 3 × 1 centered in the Gamma high symmetry point. To reach geometry optimization, the energy differences and force components must be less than $1.0 \times 10^{-6}$ eV and 0.01 eV/Å², respectively.

**Discussion:**

**Interplanar spacing, lattice constant, and diffraction peak width:**

Figure 1(a) shows a waterfall plot of the symmetrically coupled θ/2θ XRD scan covering a wide angular range. Diffraction peaks corresponding to the $Al_2O_3$ (006) and (0012) reflections were observed, while the Ta (222) peak demonstrated a (111) growth orientation. From the d-spacing of the symmetric (222) reflection, the lattice spacings as a function of film thickness were produced using the relationship of Miller indices to lattice spacing for cubic crystals, $a =$

$d\sqrt{h^2 + k^2 + l^2}$ where, d is interplanar spacing, h, k, l are the Miller indices, and $a$ is the lattice constant.

The symmetric (222) reflection is shown as a function of thickness in Figure 1(b); we do not observe any significant variation in the (111) interplanar spacing as a function of thickness; only very minor variations in lattice constant up to a maximum of 0.003 nm can be observed, although these variations are near instrumental error. Figure 1(c) shows cubic lattice constant as a function of thickness, revealing a constant of 0.331 nm in agreement with the bulk lattice constant. For some thicker films, very weak reflections of beta-Ta 313 can be observed at 58-60 degrees, indicating small inclusions of beta in otherwise very-well oriented α-Ta (111). Although the source of the β-Ta is still under investigation, we suspect it could form during surface oxidation or from residual gas in the growth system during cool-down post-growth. Additionally, the full-width at half-max (FWHM) of the α-Ta (222) peak rapidly improves from a significantly broad peak of 4.3 degrees FWHM for the thinnest films to 0.2 degrees FWHM for the thickest films with an instrumental broadening of ~0.1 degrees.

**XRR and Thin-film model:**

Three distinct layers were observed in our films: an interfacial layer of Ta-O caused by the nucleation of the Ta on the oxygen-terminated c-plane $Al_2O_3$ substrate, the α-Ta thin film layer, and an amorphous Tantalum Oxide layer generated from air exposure. Utilizing built-in XRR simulations capabilities in the Rigaku SmartLab software, a three-layer model is utilized for theoretically computing XRR profiles and quantitatively compared to XRR measurements of as produced Ta (111) films; see Figure 2 for experimental and theoretical fits. Figure 3 shows trends in film density, roughness, and thickness as a function of total film thickness. When building the thin-film model for XRR simulations, two-, three-, and four-layer models were considered. Tell-tale beat patterns within the XRR plot at approximately 1.5 degrees eliminated the potential for a 2 layers model, as this was insufficient to replicate this feature; while a 4 layer model does work but converges to the same results as a three-layer model, ergo a three-layer model prevails as to not overparameterized the theoretical modeling.

At the interface between the $Al_2O_3$ substrate and Ta layer, a very thin interfacial layer forms from the nucleation of the Ta onto the $Al_2O_3$ 001 surface. A relatively consistent 0.8 +/- 0.2 nm thick layer independent of the film thickness, see Figure 3(g). The density of this interfacial layer is also independent of thickness and just shy of the bulk density (16.6 g/cm3) of α-Ta at a value of ~15 g/cm3, see Figure 3(h). Likewise, the roughness of this interfacial layer is independent of thickness but has a roughness quantity of approximately 60% of that of the observed thickness, see Figure 3(i).

After this thin interfacial layer, a pure α-Ta layer is formed. The thickness of this layer follows a highly linear, nearly one-to-one trend with total thickness as expected, Figure 3(d). However, the density of the Ta layer is almost constant with a density of ~12.1 g/cm3, which is 27% lower than the bulk value, see Figure 3(e); previous research on sputter films found the angle of declination to affect the densification of the resultant film, where at normal incidence densification is in agreement with our findings[95]. Interestingly, when compared to the density interfacial Ta layer, a marked reduction in density is observed. While less densification of sputter films is known to occur, we anticipate this loss in density to be caused by space between columnar growth, although further work is needed to determine what impacts densification[96–98]. Compared to previous thin films of metal, a reduction of densification to ~20% is typical. The roughness of the pure Ta layer has a slight upward trend as a function of thickness but exhibits little variation when comparing films of similar thicknesses. The roughness grows from ~0.2 nm to ~1.0 nm across the thickness series.

At the surface, the tantalum-oxide layer thermally oxidizes up to a thickness of 2.4 +/- 0.177 nm and is measured with a slightly lower density than nominal $Ta_2O_5$, as seen in Figure 3(a,b). In the limit of ultra-thin films, the entirety of the Ta film oxidizes, and the thickness is constrained by the film thickness. Junsik Mun et al. found for Ta (110) thin films an amorphous oxide of up to ~3 nm thick driven by air exposure. [64]. The surface roughness of the oxide layer has a slight upward trend as a function of thickness but shows significant variation from film to film, even amongst films of similar thicknesses, see Figure 3(c). The roughness grows from ~0.2 nm to ~1.0 nm across the thickness series. Not surprisingly, the roughness variation of the oxide layer correlates well with the variation of Ta roughness, as can be seen in Figure 3(c,f). This is to be expected, as the final roughness of the Ta layer sets the overall meandering of the surface morphology for the amorphous oxide to replicate.

**Film surface morphology**

Figure 4 shows an array of AFM images of selected films covering the thickness series and beyond demonstrated in the XRR analysis. We extract the terrace width and film roughness from the AFM imaging as a function of thickness plotted in Figure 4(i). Terrace widths are extracted from a line cut analysis of several grains to compute a mean and standard deviation. Roughness calculations are performed using Gwyddion's native surface statistical analysis package. For the thickest films (250 nm), a striking triangular morphology emerges with a characteristic terrace width of 120 +/- 45 nm. As the film series thins, the terrace width shrinks, and the regularity of the triangular terraces appears rounder. We find an upward trend in terrace size as a function of thickness ranging from 20 +/- 10 nm for the thinnest films up to 120 +/- 45 nm for the thickest. Initially, the films nucleate with small islands for thickness below 30 nm, Figure 4(f-h); as the films thicken, these islands tend to grow together, forming a maze of island growth for film thickness below 175 nm but larger than 30 nm, see Figure 4(b-e), where a triangular island morphology finally emerges for thickest films of 250 nm Figure 4(a). We suspect some of the smallest islands can be triangular but are limited in resolution by AFM tip size to characterize the triangular geometry robustly. The roughness of the film as a function of thickness follows a constant trend and agrees with the trends observed in XRR with 500 +/- 75 pm roughness. Additionally, while Figure 4 demonstrates characteristics of the columnar growth associated with lower density of sputtered films, a larger study of morphological trends would be necessary to confirm the correlation. While Figure 4(b-f) appears to show smaller island growths much higher than the surface average, the restricted color ranges exaggerate this effect, and these small, bright portions are not seen as being overly morphologically significant.

**Oxide Simulation Results**

In **Error! Reference source not found.**(a-c), the oxidation process is depicted, which was carried out progressively by placing oxygen atoms in the most stable positions. These positions were determined by analyzing different configurations and identifying those with the lowest energy. Strong structural changes are induced by oxygen presence on Ta. For example, compare the difference in interplanar distances when going from the clean surface (see Figure 5(f)) to the one with the highest oxidation coverage (Figure 5(e)). The reference interplanar distance (magenta horizontal lines) is 2.8 Å. As oxidation increases to approximately ~33% of the monolayer content, the interplanar distance increases to 2.8 Å. Upon raising the oxygen content to ~66% (**Error! Reference source not found.**(b)) and 100% (Figure 5(c)), the interplanar distances increase to 3.9 Å. A sharp contrast occurs when oxygen coverage exceeds the 1:1 Ta:O ratio. **Error! Reference source not found.**(d-e) clearly evidences the formation of an amorphous phase, indicating a significant modification of the atomic structure at the surface and an increase of the interplanar distance of 7.7Å and 8.9Å for ~166% and 250% of oxygen content. Specifically, **Error! Reference source not found.**(e) represents a system with a Ta:O atomic ratio of 1:2.5. The amorphization of the surface is attributed to the strong interaction between oxygen and tantalum atoms, which induces a redistribution of the Ta atom positions and disrupts the crystalline periodicity. This phenomenon is consistent with previous studies on the oxidation of Ta 110, where oxygen incorporation has been observed to generate disordered networks due to the difference in electronegativity and atomic size between the two elements[64].

To compare to our experimental results, we plot the theoretical density of the $Ta_xO_y$ as a function of coverage (Figure 7) revealing a downward exponential relationship trending toward bulk Ta2O5 density of 8.2 g/cm$^3$. After 166% of coverage, the theoretical surface oxide forms a density that is in agreement with the bulk. As a function of Ta thickness, we find a very slight upward trend in density as a function of thickness, but on average, measure a density of ~6 g/cm3, slightly less dense than bulk.

**Transition Temperature and RRR Results**

Figure 6(a) shows the sheet resistance ($R_s$) of the films with temperature. We were able to observe a superconducting transition for all films with a thickness ($t_{total}$) of 15 nm and higher, whereas the extrapolated critical temperature ($T_c$) for the thinner films was lower than the base temperature (~2.9 K) in our system.

Figures 6(c) and 5(d) show the $T_c$, defined as the temperature where $R_s$ decreases to half of the normal state sheet resistance ($R_n$), as a function of $d$ and $R_n$. This film thickness-driven suppression of superconductivity involves interplay with the disorder. Film resistivity (and thus disorder scattering) rapidly increases at low thickness, as shown in Figure 6 (e). The combined effect of dimensionality reduction and disorder on superconductivity can be compactly expressed as:

$$dT_c = AR_S^{-B} \qquad \qquad \textit{Equation 1}$$

This empirical power law was observed to collapse $d \cdot T_c$ vs $R_s$ data onto scaling curves in many metallic superconductor materials [89]. Figure 6(b) shows that this scaling is also holds for Ta films. $R_s$ values at 5 K or room temperature can be used, giving (A, B) = (624.32, 0.84) and (118.14, 0.45), respectively. The former set of A and B coefficients has practical utility in enabling quick estimation of expected superconducting $T_c$ values from resistance measurements in ambient conditions. Across many superconducting metals, the extracted coefficients were found to loosely cluster near $B \approx 0.8 - 1.1$ and $\log_{10}(A) = 1.23 + 2.64 \cdot B$, hinting at a universal description [89]. Our measured value of $B = 0.84$ for Ta is well within the previously observed scatter across materials. The universal logarithmic scaling trendline overestimates $\log_{10}(A) = 3.45$, with the measured value being 2.80. These are subtle deviations from universality that can be speculated to be determined by granularity or crystallinity of the material as in [89], but further comparison across film thickness series with consistent and distinct morphologies is needed.

**Conclusions**

In conclusion we explored a thickness dependence of α-Ta (111) thin films ranging from ultra-thin (~2 nm) to thick (~250 nm) films grown by sputter epitaxy on c-plane sapphire substrates. A combination of X-ray reflectometry and diffractometry alongside atomic force microscopy to characterize the resultant crystal structure, morphology, substrate film interface, and amorphous oxide. Films grew with a mono-oriented α-Ta (111) films with a lattice spacing in agreement with bulk Ta value of 3.31 Å and low roughness of ~700 pm. The films were found to nucleate with small islands, whose terrace width grows linearly as a function of thickness until ~150 nm, where the terrace width levels off. We uncover a pseudomorphic layer with a critical thickness of ~1 nm and a self-limiting amorphous oxide that grows to a thickness of ~2.25 nm, with both of these layers independent of thickness beyond ~4 nm total thickness. Our quantum mechanical computations demonstrate that the oxidation of the α-Ta (111) surface with a Ta:O ratio between 1:0 and 1:1 induces structural changes. As the oxygen content increases within this range, Ta atoms undergo contractions or expansions in their interatomic distances. However, when the Ta:O ratio exceeds 1:1, the crystalline structure undergoes significant deformation, leading to the formation of an amorphous TaO phase. Our computational results highlight how the progressive incorporation of oxygen alters the structural stability of tantalum, which has substantial implications in applications where oxidation reduces the material performance.

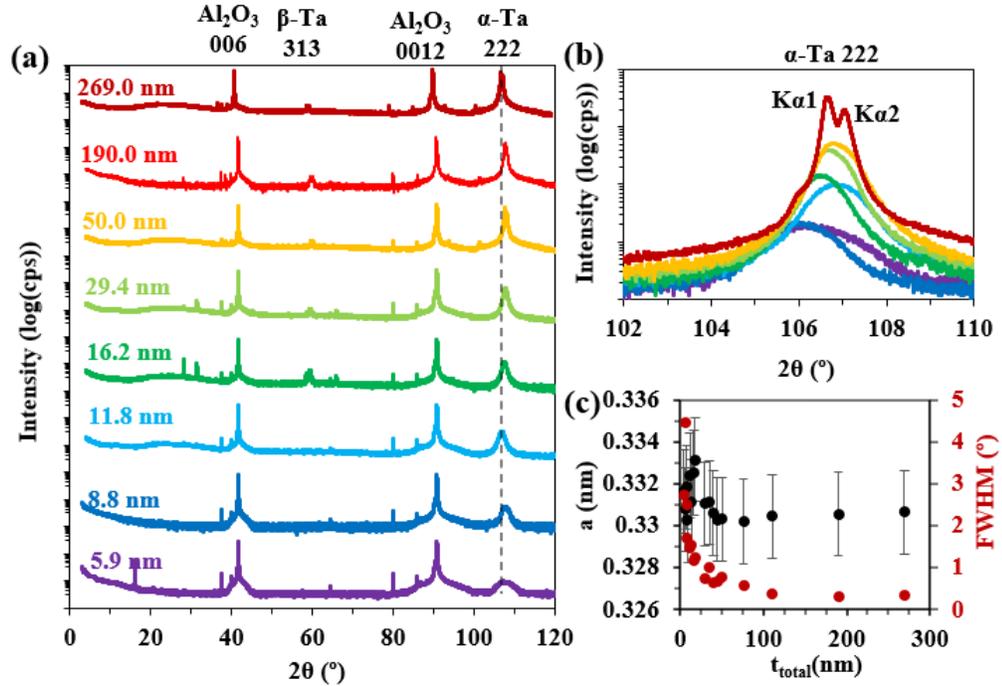

Figure 1 X-ray diffraction measurements and diffraction line analysis. (a) Symmetric coupled-scans of θ-2θ spanning the thicknesses studied. A pronounced α-Ta(222) peak is observed and indicated by the black dashed line. A weak β-Ta can be observed on occasion. (b) Zoom in of the α-Ta(222) as the same function of thickness in (a) demonstrating the alignment shift as function of thickness and full-width at half-max decrease with increasing thickness. (c) X-ray diffraction line analysis of the α-Ta(222) peak. The lattice constant is extracted from the (222) position at each thickness and plotted in black circles. The full-width at half-max of the Kα1 is extracted from fitting the Kα1 Kα2 X-ray lines to the observed peaks and plotted in red circles as a function of thickness.

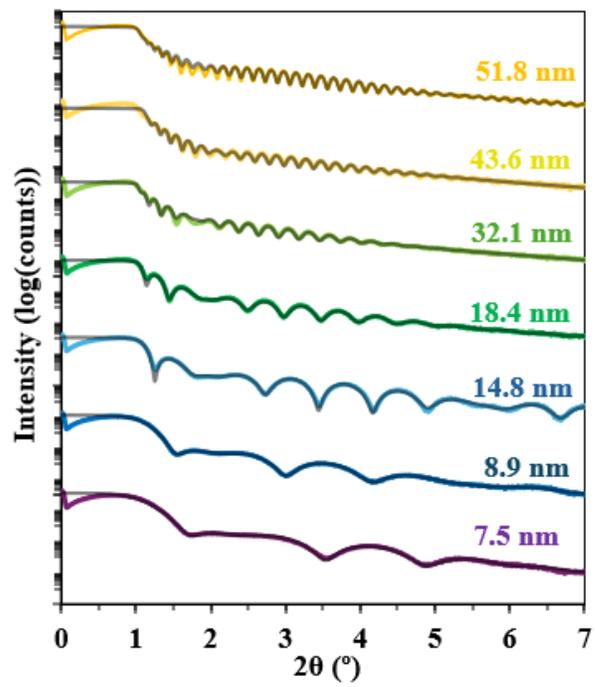

Figure 2 X-ray reflectometry measurements and best theoretical fits for α-Ta films as a function of thickness. Experimental data is shown in black, while theoretical fit is shown in red.

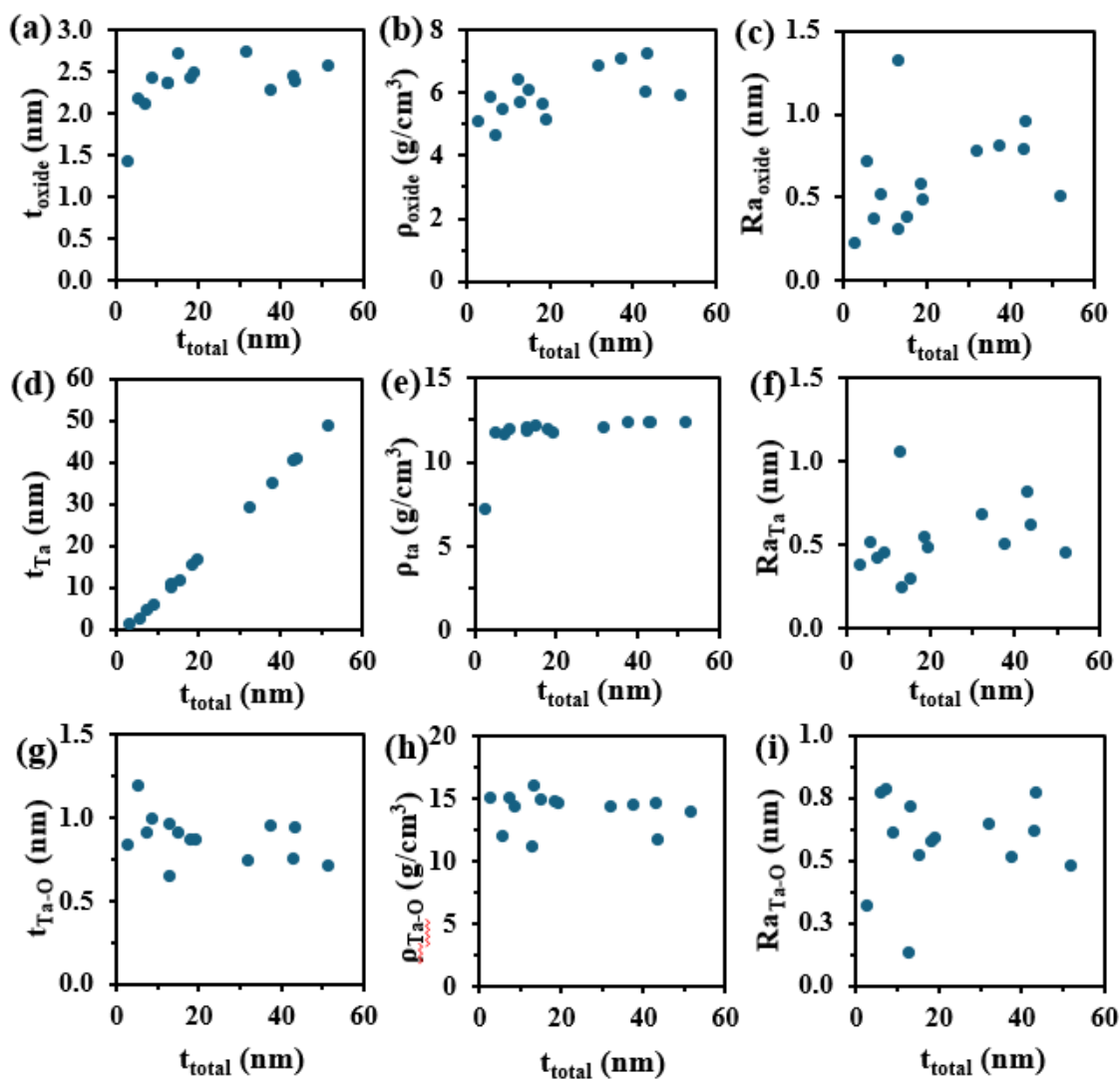

Figure 3 Extracted quantities from best fit to X-ray reflectometry data for each layer in the fitting model. (a-c) corresponds to the native oxide layer computing thickness, density, and roughness as a function of total thickness. (d-f) correspond to the Ta layer computing thickness, density, and roughness as a function of total thickness. (g-i) correspond to the substrate-interface layer computing thickness, density, and roughness as a function of total thickness.

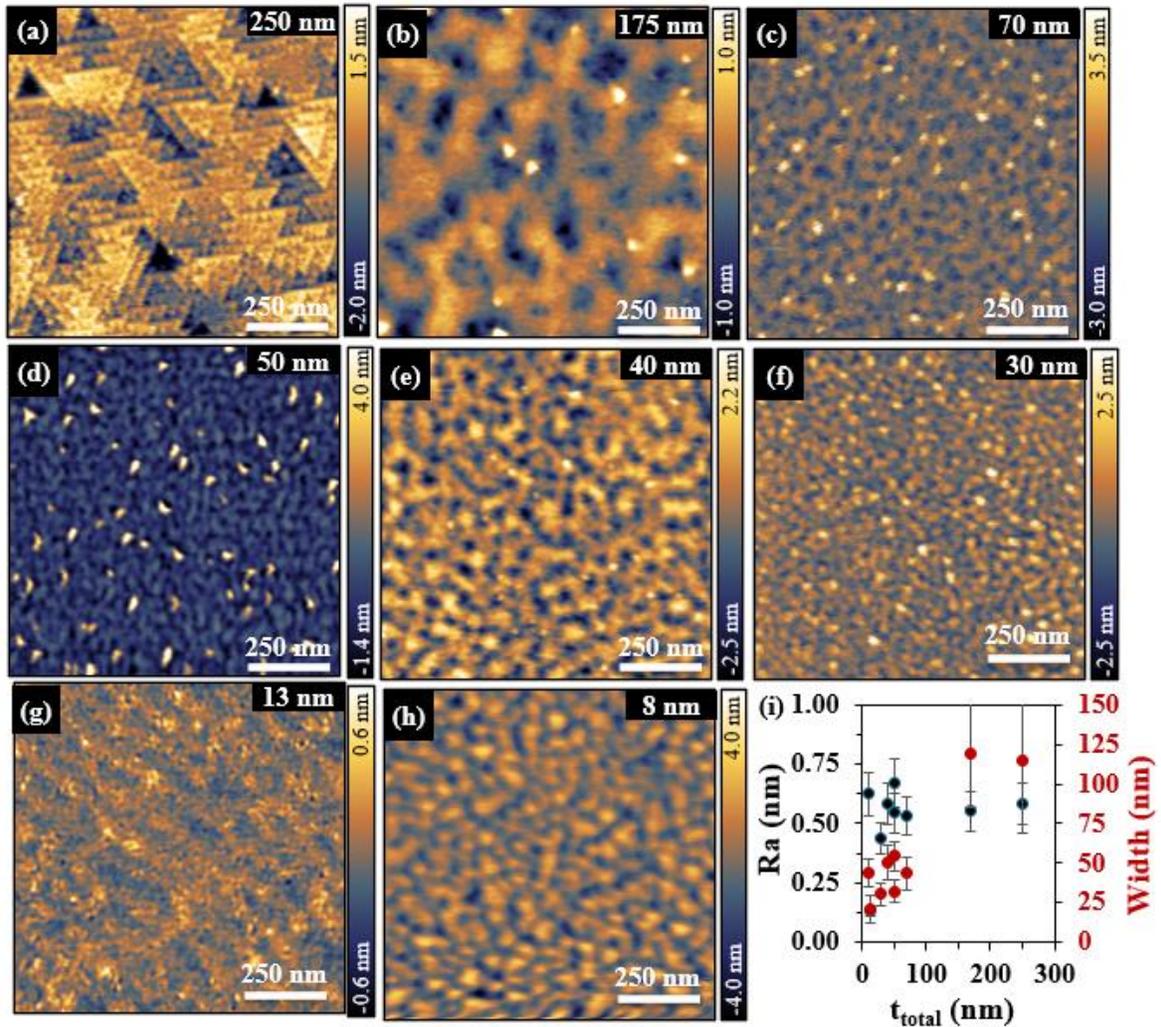

Figure 4 Atomic force micrographs and extracted roughness and terrace width of a series of thicknesses. Each panel (a-h) correspondes to a different film thickness. While (i) shows the computed roughness and terrace width.

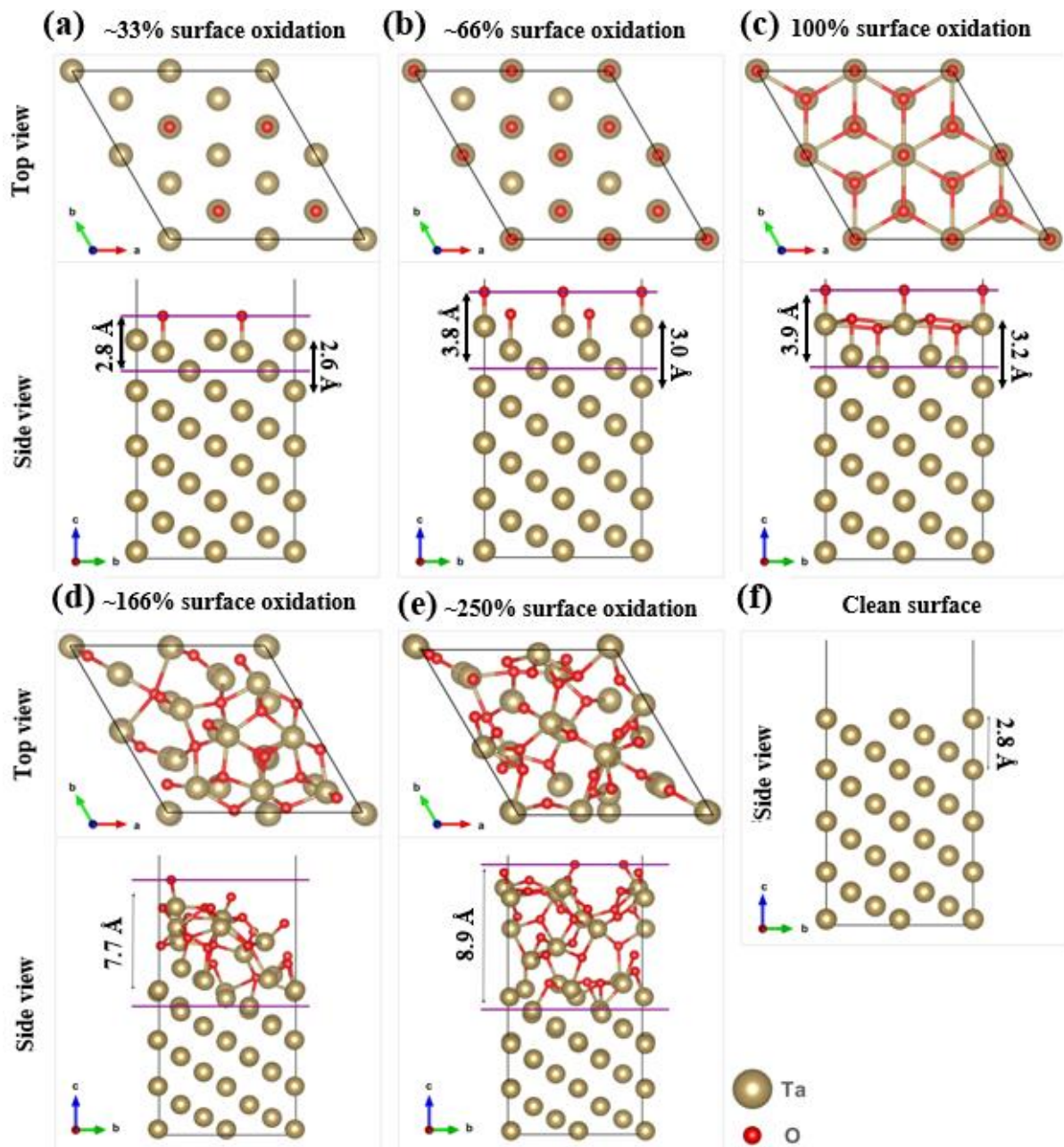

*Figure 5* α-Ta Oxidation process with increasing coverage from (a) to (e). As oxidation increases, surface deformation is induced (see horizontal magenta lines in each oxidation coverage to notice the distortion). In (d) and (e), the approximate thickness is 7.7 Å and 8.9 Å, respectively. (f) depicts the clean surface before any oxidation

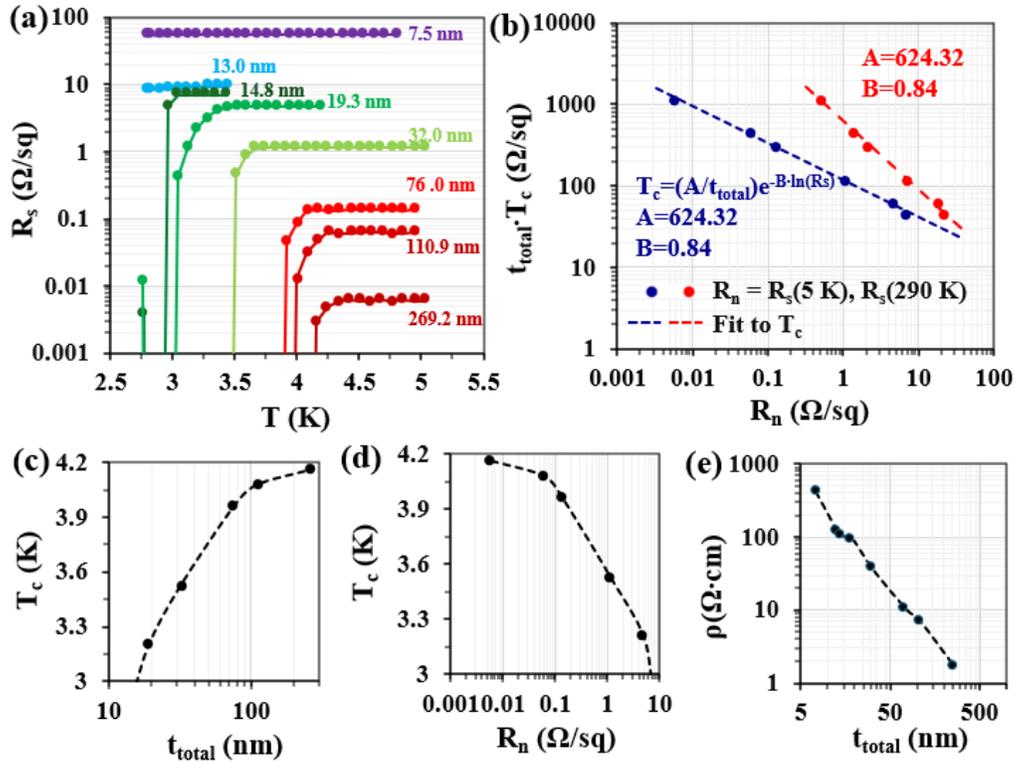

Figure 6 Superconducting transitions and scaling with thin film properties. (A) Sheet resistance as a function of temperature for all films. (B) Scaling fit to eq. (1) using measured sheet resistance (at 5 or 290K), critical temperature, and total film thickness. (C), (D), (E) Scaling of critical temperature and resistivity with total film thickness and sheet resistance.

**SUPPLMENTAL**

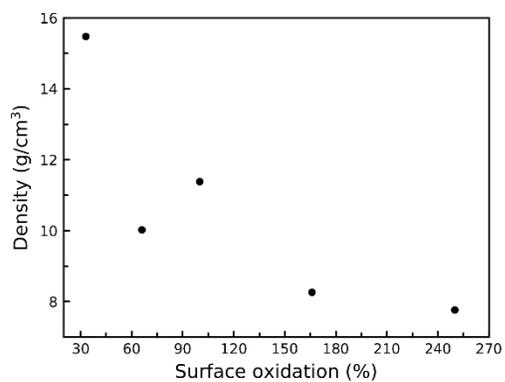

Figure 7. Density of Ta surface oxide computed from DFT as a function of coverage showing a downward trend from nearly the Ta bulk density toward the bulk Ta2O5 density.